# Behavioral similarity of dissipative solitons in an ultrafast fiber laser

Ying Yu,[1] Cihang Kong,[1] Bowen Li,[1] Jiqiang Kang,[1] Yu-Xuan Ren,[1] Zhi-Chao Luo,[1,2,3]
and Kenneth K. Y. Wong[1,4]

[1]Department of Electrical and Electronic Engineering, The University of Hong Kong, Pokfulam Road, Hong Kong, China

[2]Guangdong Provincial Key Laboratory of Nanophotonic Functional Materials and Devices & Guangzhou Key Laboratory for Special Fiber Photonic Devices and Applications, South China Normal University, Guangzhou 510006, China

[3]e-mail: zcluo@scnu.edu.cn

[4]e-mail: kywong@eee.hku.hk

**Abstract:** Dissipative solitons are non-dispersive localized waves that exhibit intriguing dynamics. In optical systems, the discovery of new soliton dynamics and how to accurately manipulate dissipative solitons are still largely unexplored and challenging in the ultrafast fiber laser community. Herein, we unveil a new type of dissipative soliton behavior in a net-normal-dispersion bidirectional ultrafast fiber laser. That is, the bidirectional dissipative solitons will always reveal similar spectral and temporal characteristics through common gain and loss modulation, even if the transient instability is involved. The behavioral similarity enables us to accurately design the soliton patterns by introducing seed pulses through loss modulation. As a proof-of-concept application, the precise and flexible manipulation of multi-soliton patterns is demonstrated. These findings will shed new insights into the complex dissipative soliton dynamics, and also beneficial for designing ultrafast lasers with desirable soliton patterns for practical applications.

Solitons, as remarkably localized structures in nonlinear systems, are discovered in diverse fields ranging from fluid dynamics to plasma physics to nonlinear optics [1-4]. Subsequently, the physicists realized that the nonlinear dissipative systems could support solitary waves in a wide range of parameters as well. Thus, the concept of soliton has been gradually extended to dissipative soliton [5-7]. In addition to the balance between nonlinearity and dispersion/diffraction, the existence and stability of dissipative solitons are highly dependent on gain and loss [6]. Relying on the framework of dissipative systems, a rich landscape of dissipative soliton dynamics, which had not been observed in conservative systems, were unveiled [8-10]. In fact, one pronounced feature of nonlinear dissipative systems is that a given set of system parameters, rather than the initial conditions, would predetermine the dissipative soliton parameters such as energy and profile [10]. Moreover, the perturbation of an optical field could be treated as the seed signal to the evolution of the dissipative soliton. These imply that in the same dissipative system, any similar perturbations of gain and loss in two independent optical fields will lead to analogous soliton behaviors. Therefore, it is possible to generate two or more dissipative solitons with similar behaviors or parameters through preset perturbations, which would be of great interest for the fundamental study of nonlinear soliton dynamics. On the other hand, in a given dissipative system, as the final profile of dissipative solitons after nonlinear evolution is not stringent with the seed signals, when the system parameters are optimized for multi-soliton propagation, it could be possible to accurately design and manipulate multiple solitons with desirable temporal patterns by setting perturbation positions. Note that the manipulation of multi-soliton patterns is important to its applications in fields such as laser material processing [11] and soliton-encoded bit stream for communications [12]. Thus, developing a new method to control multi-soliton patterns would expand the application horizon of dissipative solitons.

The mode-locked ultrafast fiber lasers, which are essentially dissipative systems, have been demonstrated as an excellent test bed in exploring dissipative soliton dynamics [13-18]. Generally, the ultrafast fiber lasers operate in unidirectional and single wavelength regime, meaning that only "single" optical field propagates in the laser cavity. Therefore, similar gain or loss modulation/perturbation between two independent optical fields is difficult to realize in the conventional ultrafast fiber lasers. Fortunately, recent results demonstrated that the ultrafast fiber laser could be mode locked in a bidirectional manner [19-21]. In this case, the lightwaves propagate in two opposite directions in the same laser cavity, indicating that they share the same gain medium, saturable absorber and net dispersion. Therefore, the bidirectional ultrafast fiber lasers possess inherent advantages of common gain and loss modulation between two optical fields in a single dissipative system owing to the unique geometrical structure. So far, only few reported bidirectional mode-locked fiber lasers are operating in normal-dispersion regime, which are good platforms for dissipative solitons, and rare dynamic analysis was involved. As a consequence, it would be interesting to identify whether the dynamics of bidirectional dissipative solitons exhibit analogous behavior in a normal-dispersion fiber laser or not, and further explore potential applications of this phenomenon.

Here, we report the dynamics of dissipative solitons in a CNT mode-locked bidirectional fiber laser operating in normal-dispersion regime. Benefitting from the powerful dispersive Fourier transform (DFT) technique [22,23], shot-to-shot spectra with sub-nanometer resolution are captured. It is found that the dissipative solitons from clockwise (CW) and counter-clockwise (CCW) directions always appear in pairs and have very similar parameters such as repetition rate, spectral bandwidth and profile, owing to the common gain/loss modulation. In particular, during dynamic events they also follow the same evolution trend. By utilizing this behavioral similarity, the temporal patterns of the generated dissipative solitons could be flexibly controlled by adjusting or setting the perturbation points, i.e., CNT positions in the laser cavity. Our findings demonstrate the behavioral similarity of dissipative solitons in ultrafast laser systems, which will be meaningful to the communities dealing with the fundamentals of solitons and laser technologies.

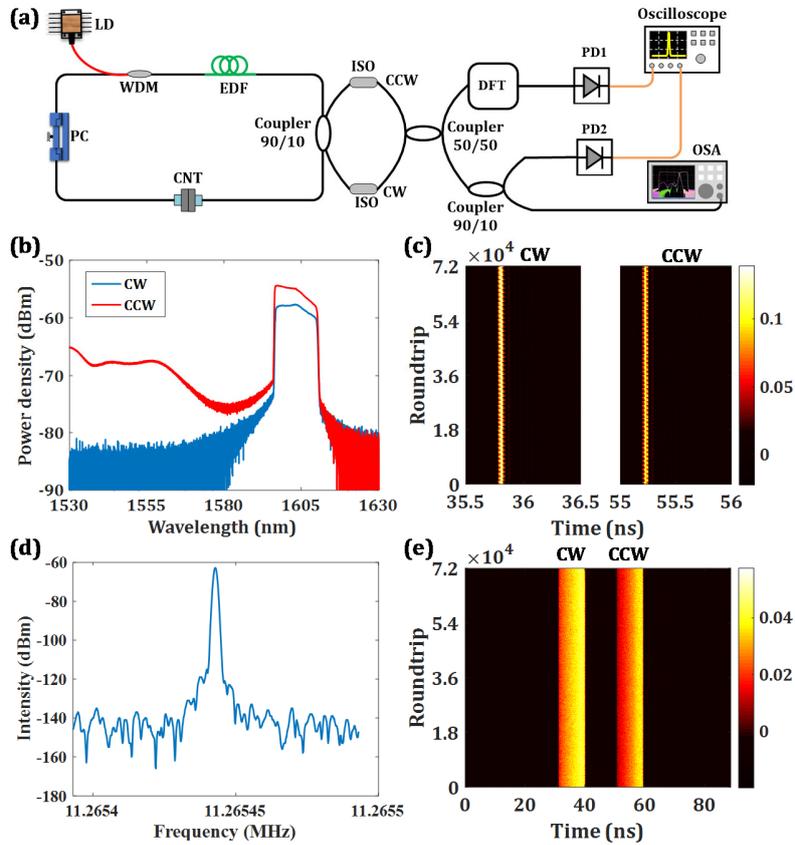

Fig. 1. (a) Schematic illustration of the laser cavity. (b)-(e) Experimental measurement results when the laser operates in a stable mode-locking state. (b) Average spectra recorded with an OSA. (c) Temporal evolution. (d) RF spectrum. (e) Spectral evolution.

Figure 1(a) shows the schematic of the fiber laser and measurement setup. The fiber laser was a net-normal dispersion mode-locked fiber laser based on a carbon nanotube (CNT). The gain fiber was a piece of 13.5-m lowly erbium-doped fiber (EDF). A polarization controller (PC) was inserted into the cavity to optimize the polarization state. Since no isolator (ISO) was placed in the cavity, the lightwave could propagate in both CW and CCW directions. A 90/10 coupler was used to extract 10% of the power propagating in each direction for measurement. The total cavity length was around 18 m, corresponding to a fundamental repetition rate of 11.3 MHz.

In order to accurately compare the behavior between opposite directions, the CW and CCW outputs were combined by a 50/50 coupler with proper delay to avoid temporal overlap and measured by the same devices. The temporal information was detected by a 32-GHz photodiode (PD2, HP 83440D) and digitized by a 20-GHz real-time oscilloscope (Lecroy SDA 820Zi-B); while the spectral information was detected by an optical spectral analyzer (OSA, Agilent 86142B) and DFT technique simultaneously. The OSA provided average spectrum, while the DFT branch, which was comprised of a spool of dispersion-compensating fiber (DCF) with -577 ps/nm dispersion and detected by a 12-GHz photodiode (PD1, New Focus 1544-B), provided the single-shot real-time spectrum with 0.17-nm spectral resolution [24].

Depending on the pump levels, the laser could operate in different states. When the pump power was ~16 mW, the laser operated in a stable single-soliton state. Surprisingly, the two directions could always be mode-locked simultaneously and revealed similar characteristics. Figure 1(b) shows the spectra captured by the OSA. Both spectra cover 15-nm bandwidth with a central wavelength of 1603.5 nm. The difference is CCW direction has higher pulse energy and amplified spontaneous emission (ASE) noise. The large ASE noise in CCW direction was originated by backward pumping of the EDF. Figure 1(c) shows the temporal evolution of both directions over ~72,000 roundtrips. The separation between CW and CCW solitons remained unchanged at 19.43 ns, which means that they had the same repetition rate. The radio frequency (RF) spectrum of the combined signal was detected by an electrical spectral analyzer (ESA) with 1-Hz resolution. There is only one fundamental frequency at 11.3 MHz shown in Fig. 1(d), which further confirms the synchronization. It should be also noted that the solitons propagate in opposite directions collide within the cavity in every roundtrip. According to the cavity length, the output fiber length and the measured separation, it can be calculated that one collision point inside the cavity is the location of CNT. The coupling between two counter propagating waves in the CNT could lead to a lock-in effect and the lock-in frequency range is related to the pulse width [25]. In this experiment, a small deviation in the repetition rate may be induced by the asymmetric structure of the laser cavity. However, the results reveal good temporal synchronization in the whole process, which may be attributed to the collision in CNT and the large pulse width in normal-dispersion mode-locking. In addition, the single-shot spectra in Fig. 1(e) show that the dissipative solitons from both directions were highly stable. Note that dissipative soliton parameters in nonlinear dissipative systems would be predetermined by the system parameters. Moreover, in this laser cavity pulses propagating along opposite directions share the same net dispersion, nonlinearity and net gain when they operate in stable mode-locking state. The collision in the CNT means that the solitons from different directions pass through the CNT simultaneously, which will lead to equivalent loss modulation for their common pass. Thus, this bidirectional laser system determines the similar performance of dissipative solitons at two directions, such as repetition rate and spectral bandwidth.

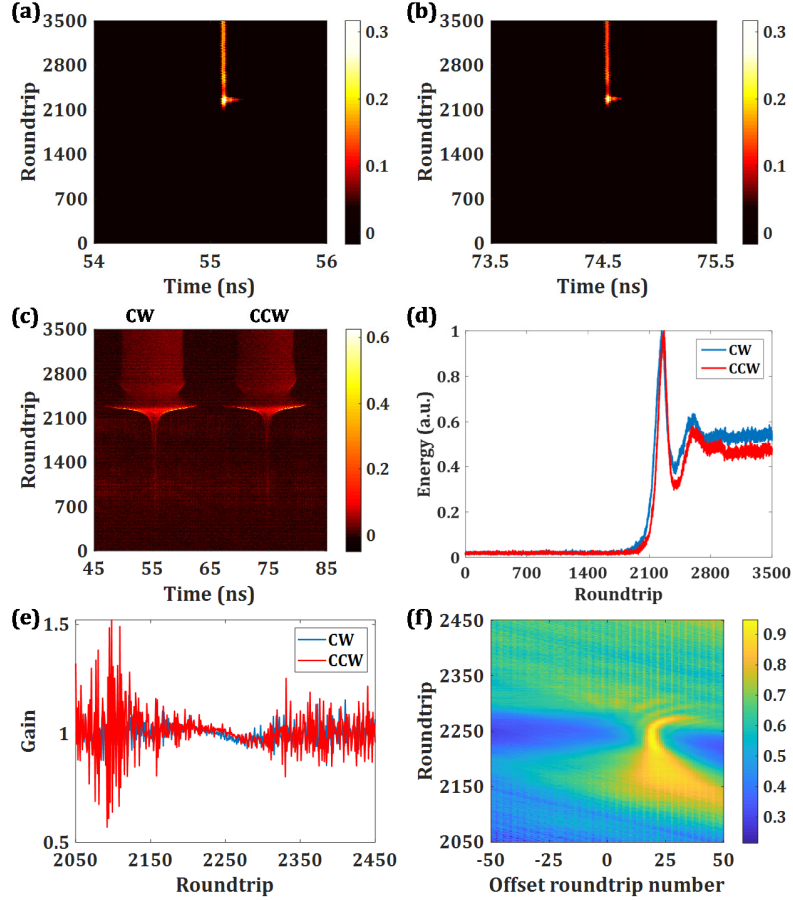

Fig. 2. Build-up of mode-locking. Temporal evolution in (a) CW and (b) CCW directions. (c) Shot-to-shot spectral evolution. (d) Energy evolution. (e) Net gain evolution. (f) Spectral cross-correlation map.

To further explore the features of behavioral similarity, the build-up process of mode-locking was observed and the results are shown in Fig. 2. Figure 2(a) and (b) show the temporal evolution along CW and CCW directions, respectively. Note that once the seed pulse is established in one direction, another seed pulse will be generated in the opposite direction owing to the loss modulation of CNT. Even though the solitons varied dramatically during the build-up process, the temporal separation between CW and CCW directions could be still maintained. They also experience similar spectral evolutions, from narrow-band noise pulse, energy enhancement, spectral broadening to stable mode-locking, as shown in Fig. 2(c). Figure 2(d) demonstrate the individually normalized energy evolution of two directions based on the peak power in Fig. 2 (a) and (b). Their energy evolution follows the same trend with a delay of ~20 roundtrips. The net gain of every roundtrip is calculated and shown in Fig. 2(e). Considering the large intensity fluctuation during the whole process and the limited saturation power of the detectors, the measured signal-to-noise ratio (SNR) of stable mode-locking state has to be compromised. Meanwhile, the starting phase of build-up process has small signal intensity. Therefore, the calculated net gain before 2150th roundtrip and after 2300th roundtrip is greatly affected by the detection noise and shows strong fluctuation. However, the net gain evolution during 2150th and 2300th roundtrip, which is the main part of the whole build-up process, clearly show that the net gains of different directions follow the same evolution trend with a slight difference. In order to compare their spectral behavior in detail, the cross correlation between single-shot spectra from opposite directions was computed. For every roundtrip starting from 2050th to 2450th propagates in the CW direction, its spectral cross correlation with 100 successive roundtrips around in the CCW direction are calculated. In Fig. 2(f), each value represents the cross correlation of the two single-shot spectra with no delay and reveals their similarity. Actually, because of the modulation effect of CNT, the build-up process of both directions always starts simultaneously. After generation at the CNT together, the solitons from opposite directions will collide at the CNT in each roundtrip and share same loss modulation. Note that the CW and CCW solitons also share the same gain fiber. That is why their net gain follow the same evolution trend. For every roundtrip from CW direction, there is always one roundtrip from CCW direction which has a very high spectral correlation with it, indicating that they have highly spectral similarity. However, the different pumping directions make the gain of different directions slightly different, as observed in Fig. 2(e), and further leading to the bending in the spectral cross-correlation map.

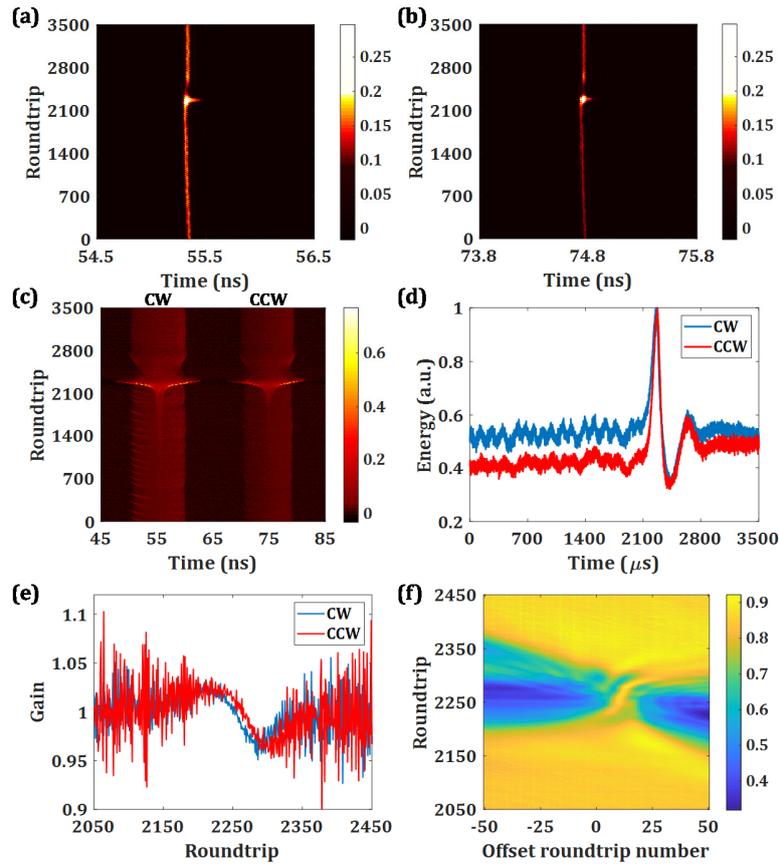

Fig. 3. Experimentally measured dynamic event. Temporal evolution in (a) CW and (b) CCW directions. (c) Shot-to-shot spectral evolution. (d) Energy evolution. (e) Net gain evolution. (f) Spectral cross-correlation map.

In addition to the build-up process, another special dynamic event has also been observed and analyzed, as shown in Fig. 3. In this event, the solitons operate in quasi-stable states initially. However, there is obvious spectral vibration in the CW direction for longer wavelength [left side of CW spectra in Fig. 3(c)], which leads to the energy fluctuation depicted in Fig. 3(d). At ~2100th roundtrip, the soliton runs into an unstable state that is quite similar to the main part of build-up process, but with a broad-band spectral pedestal, and finally evolves into stable mode-locking state. During the whole process, the temporal separation between CW and CCW directions remains stable, even though the repetition rate changes slightly, as shown in Fig. 3(a) and (b). Their net gains which were presented in Fig. 3(e) still maintain the same evolution trend, which ensures the highly spectral similarity presented in Fig. 3(f). Again, the bending comes from minor gain difference, as revealed in Fig. 3(e). At the beginning, the spectral vibration of CCW direction is not as strong as CW one and not shown clearly in Fig. 3(c). However, it could be confirmed from the temporal synchronization [Fig. 3(a) and 3(b)] and energy fluctuation [Fig. 3(d)]. It should be noted that, even though the vibration intensity may vary, the spectral and temporal evolution along the two directions follows the same trend because of the similar gain and loss modulation.

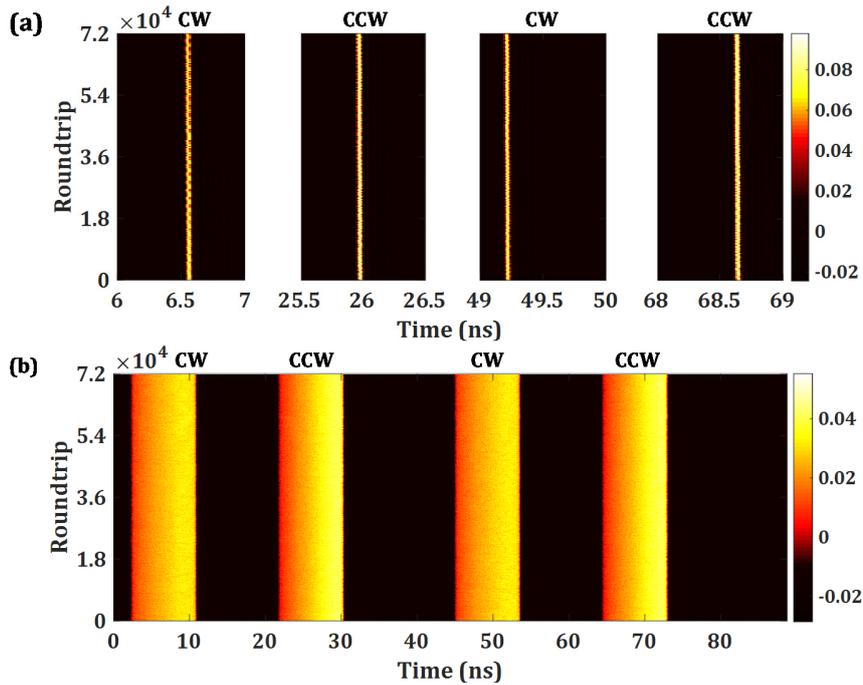

Fig. 4. Experimentally measured dual-soliton state. (a) Temporal evolution. (b) Spectral evolution.

It has already been demonstrated that similar gain and loss in one dissipative system result in highly spectral and temporal similarity in single-soliton regime, no matter in stable state or transient state. As it is well known that fiber lasers could run into more complex multi-soliton regime by controlling the pump power. Therefore, it is natural to ask whether the behavioral similarity could exist in multi-soliton regime. To this end, the pump power was further increased to ~19 mW, and the laser operation switched to a dual-soliton regime. Figure 4 reveals that in the dual-soliton state, CW and CCW directions still synchronize well with each other. Both directions operate in a dual-soliton state with 42.66 ns temporal separation and the delay between CW and CCW directions remains the same.

We have proved that the behavioral similarity exists in both single-soliton and multi-soliton regimes, and the solitons from different directions always arise in pairs and collide in the CNT. When the pump power is high enough, the laser could operate in multi-soliton regime. For example, in the case shown in Fig. 4, there are two independent soliton pairs with behavioral similarity. However, the temporal separation between them is generated randomly, which is unfavorable for precise control. Considering that the insertion of one more CNT could support more soliton pairs and break the "independent" relationship, it should be possible to manipulate the multiple solitons with desirable pattern. Based on this hypothesis, another bidirectional mode-locked laser with similar structure but 2 CNTs was constructed, as shown in Fig. 5(a). The fiber length $L$ between these 2 CNTs was ~1 m. When the pump power was ~16 mW, both directions operated in the single-soliton regime, and CNT1 is one collision point inside the cavity according to the calculation. Note that if the pump power level could support more solitons, CW soliton could assist generating one more seed pulse in CCW direction when it passes through CNT2 owing to the loss modulation effect of CNT. That will lead to the formation of a new soliton pair whose collision point is CNT2. The travel distance of CW soliton from CNT1 to CNT2 is $L$. Meanwhile, the CCW soliton also travels $L$. Therefore, the separation between the original CCW soliton and the newly generated CCW soliton should be 2$L$. Actually, when the pump power is increased to ~21 mW, both directions turned into dual-soliton regimes with 9.6-ns separation, which matches well with 2$Ln/c$, where $n$ is refractive index and $c$ is speed of light. The patterns from two directions are synchronized, and the results of CW direction are shown in Fig. 5(b) as an illustration. Similarly, by increasing the pump power, the newly generated CCW soliton could help generate one more seed pulse in CW direction when it passes through CNT1. Different directions help generate new seed pulses alternately, however, the temporal separation between the last two generated solitons is always determined by 2$Ln/c$. As demonstrated in Fig. 5(b), a tri-soliton pattern and a quad-soliton pattern with 9.6-ns equal interval could also been observed with a pump power of ~26 mW and ~32 mW respectively. As the equal interval in multi-soliton pattern is determined by this 2-CNT structure, it could also be possible to flexibly manipulate the interval by inserting a delay line between these 2 CNTs. Figure 5(c) shows a dual-soliton example, in which different colors correspond to different delay amounts. When the delay line was tuned with a step of 50 ps, the soliton interval of CW direction was changed by a step of 100 ps. Importantly, this structure can support fast tuning requirement because there is no mode-locking rebuilt process during the tuning.

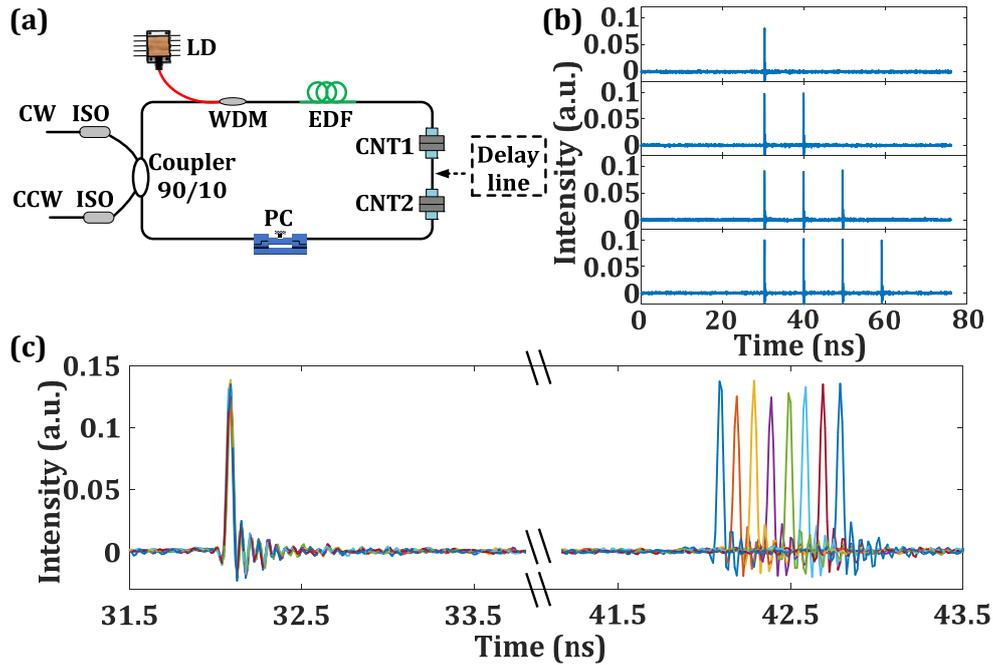

Fig. 5. (a) Experimental setup. (b) Multi-soliton pattern from CW direction. (c) Separation tunability.

In this CNT based bidirectional mode-locked fiber laser, different directions have the same net dispersion. Meanwhile, they pass through the CNT together and all the devices and splicing points are directionless, so they have similar cavity loss. Even though CW and CCW directions have different pumping structures, as they share the same piece of EDF, they will have the same gain filtering effect and similar gain. The temporal synchronization and high spectral similarity in the experimental results prove that the soliton evolution in a dissipative system can be controlled and imitated by gain and loss manipulation. In our work, only one net dispersion value is applied to investigate the behavioral similarity. Therefore, it is also worthwhile to find out whether the net normal dispersion amount could affect the behavior in the future. In addition, more complex multi-soliton pattern manipulation could also be attempted by adding more CNTs inside the laser cavity and finely design their separation.

In conclusion, we have revealed the behavioral similarity of dissipative solitons in a CNT-based bidirectional ultrafast fiber laser operating in net-normal dispersion regime. The real-time dynamics of dissipative solitons were visualized by DFT technique. Relying on the behavioral similarity of dissipative solitons, the precise and flexible manipulation of multiple solitons with expectable patterns was realized. We anticipate that the reported results will shed new light on the various communities interested in soliton dynamics and ultrafast laser technologies.

## Funding

Research Grants Council of the Hong Kong Special Administrative Region, China (HKU 17209018, E-HKU701/17, CityU T42-103/16-N, HKU C7047-16G, and HKU 17205215) and Natural Science Foundation of China (N_HKU712/16, 11874018).

# References

1. E. A. Kuznetsov, A. M. Rubenchik, V. E. Zakharov, "Soliton stability in plasmas and Hydrodynamics," Phys. Rep. 142(3), 103-165 (1986).

2. N. Akhmediev, and A. Ankiewicz, Solitons: nonlinear pulses and beams (Chapman and Hall, 1997).

3. Y. S. Kivshar, and G. P. Agrawal, Optical Solitons: From Fibers to Photonics Crystals (Academic, New York, 2003).

4. T. Dauxois and M. Peyrard, Physics of Solitons (Cambridge University Press, Cambridge, England, 2006).

5. E. Picholle, C. Montes, C. Leycuras, O. Legrand, and J. Botineau, "Observation of dissipative superluminous solitons in a Brillouin fiber ring laser," Phys. Rev. Lett. 66(11), 1454-1457 (1991).

6. N. Akhmediev and A. Ankiewicz, editors, Dissipative Solitons (Springer, Berlin, 2005).

7. F. W. Wise, A. Chong, and W. H. Renninger, "High-energy femtosecond fiber lasers based on pulse propagation at normal dispersion," Laser & Photo. Rev. 2(1-2), 58-73 (2008).

8. J. M. Soto-Crespo, N. N. Akhmediev, V. V. Afanasjev, and S. Wabnitz, "Pulse solutions of the cubic-quintic complex Ginzburg–Landau equation in the case of normal dispersion," Phys. Rev. E 55(4), 4783-4796 (1997).

9. J. M. Soto-Crespo, N. Akhmediev, and A. Ankiewicz, "Pulsating, Creeping, and Erupting Solitons in Dissipative Systems," Phys. Rev. Lett. 85(14), 2937-2940 (2000).

10. P. Grelu, and N. Akhmediev, "Dissipative solitons for mode-locked lasers," Nat. Photonics 6(2), 84-92 (2012).

11. C. Kerse, H. Kalaycıoğlu, P. Elahi, B. Çetin, D. K. Kesim, Ö. Akçaalan, S. Yavaş, M. D. Aşık, B. Öktem, H. Hoogland, R. Holzwarth, and, F. Ö. Ilday, "Ablation-cooled material removal with ultrafast bursts of pulses," Nature 537(7618), 84-88 (2016).

12. M. Stratmann, T. Pagel, and F. Mitschke, "Experimental Observation of Temporal Soliton Molecules," Phys. Rev. Lett. 95(14), 143902 (2005).

13. D. Y. Tang, H. Zhang, L. M. Zhao, and X. Wu, "Observation of High-Order Polarization-Locked Vector Solitons in a Fiber Laser," Phys. Rev. Lett. 101(15), 153904 (2008).

14. S. K. Turitsyn, B. G. Bale, and M. P. Fedoruk, "Dispersion-managed solitons in fibre systems and lasers," Phys. Rep. 521(4), 135–203 (2012).

15. A. F. J. Runge, N. G. R. Broderick, and M. Erkintalo, "Observation of soliton explosions in a passively mode-locked fiber laser," Optica 2(1), 36-39 (2015).

16. C. Bao, W. Chang, C. Yang, N. Akhmediev, and S. T. Cundiff, "Observation of Coexisting Dissipative Solitons in a Mode-Locked Fiber Laser," Phys. Rev. Lett. 115(25), 253903 (2015).

17. K.Krupa, K. Nithyanandan, and P. Grelu, "Vector dynamics of incoherent dissipative optical solitons," Optica 4(10), 1239-1244 (2017).
18. X. Liu, X. Yao, and Y. Cui, "Real-Time Observation of the Buildup of Soliton Molecules," Phys. Rev. Lett. 121(2), 023905 (2018).
19. K. Kieu and M. Mansuripur, "All-fiber bidirectional passively mode-locked ring laser," Opt. Lett. 33(1), 64-66 (2008).
20. S. Mehravar, R. A. Norwood, N. Peyghambarian, and K. Kieu, "Real-time dual-comb spectroscopy with a free-running bidirectionally mode-locked fiber laser," Appl. Phys. Lett. 108(23), 231104 (2016).
21. X. Zhao, Z. Zheng, Y. Liu, G. Hu, and J. Liu, "Dual-Wavelength, Bidirectional Single-Wall Carbon Nanotube Mode-locked Fiber Laser", IEEE Photon. Technol. Lett. 26(17), 1722-1725 (2014).
22. Y. C. Tong, L. Y. Chang, and H. K. Tsang, "Fibre dispersion or pulse spectrum measurement using a sampling oscilloscope", Electron. Lett. 33(11), 983-985 (1997).
23. K. Goda and B. Jalali, "Dispersive Fourier transformation for fast continuous single-shot measurements," Nat. Photonics 7(2), 102-112 (2013).
24. K. K. Tsia, K. Goda, D. Capewell, and B. Jalali, "Performance of serial time-encoded amplified microscope," Opt. Express 18(10), 10016-10028 (2010).
25. M. L. Dennis, J. M. Diels, and M. Lai, "Femtosecond ring dye laser: a potential new laser gyro," Opt. Lett. 16(7), 529-531 (1991).